\documentstyle[preprint,epsf,aps]{revtex}
\begin{document}
\title{Imperfect nesting and Peierls instability for a two-dimensional 
tight-binding model}
\author{Qingshan Yuan$^{1,2}$, Tamara Nunner$^1$ and Thilo Kopp$^1$}
\address{$^1$ Experimentalphysik VI, Universit\"at Augsburg, 
 86135 Augsburg, Germany\\
$^2$Pohl Institute of Solid State Physics, Tongji University, 
Shanghai 200092, P.R.China}
\maketitle
\begin{abstract}
Based on a half-filled two-dimensional tight-binding model with 
nearest-neighbour 
and next nearest-neighbour hopping the effect of imperfect Fermi surface 
nesting on the Peierls instability is studied at zero temperature. 
Two dimerization patterns corresponding to a phonon
vector $(\pi, \pi)$ are considered. It is found that 
the Peierls instability will be suppressed with an increase of 
next nearest-neighbour hopping which characterizes the nesting deviation. 
First and second order transitions to a homogeneous state are possible. 
The competition between the two dimerized states is discussed.
\end{abstract}



\section{Introduction}

The low dimensional electronic materials are known to be very susceptible 
to a Peierls instability towards a charge density wave (CDW) state
driven by the electron-phonon interaction \cite{Gruner}. 
The presence of a lattice distortion
is usually favorable to lower the electronic energy and once this reduction 
overcomes the increase of lattice deformation energy the Peierls transition 
takes place. It has been extensively studied in a lot of quasi-one 
dimensional (1D) materials such as organic conjugated polymers (CH)$_x$ 
\cite{Farges} or inorganic blue bronzes A$_{0.3}$MoO$_3$ (A=K, Rb, Tl) 
\cite{Schlenk}, as well as quasi-two dimensional (2D) materials such as 
purple bronzes AMo$_6$O$_{17}$ (A=Na, K, Tl) \cite{Schlenk,Dumas,Qin} and 
monophosphate tungsten bronzes (PO$_2$)$_4$(WO$_3$)$_{2m}$ ($4\le m\le 14$) 
\cite{Schlenker,Wang}.

It is believed that in the Peierls transition the structure of the 
Fermi surface (FS) plays an essential role. In ideal 1D systems, the FS, being 
composed of two points seperated by $2k_F$ (Fermi wave vector), is always
perfectly nested. The lattice distortion opens a gap at the Fermi level
with the consequence that the energy gain from the electronic energy is 
always dominant, so that the Peierls instability with a metal-insulator 
transition always takes place 
(if quantum effects of phonons are not considered \cite{Zheng}).

The situation becomes richer in two dimensions because of a more complex FS 
structure. In general, the FS is not nested, i.e., a single mode of lattice 
distortion can connect only two points 
in the FS, and the gain of electronic energy from this distortion is not 
enough to overcome the increase of lattice energy. However, in some special 
cases the FS is still nested and the electronic energy may be 
lowered substantially by the lattice distortion even if a gap may be not 
fully opened at the FS as in one dimension. The simplest realization is
the 2D square lattice tight-binding model with only nearest-neighbour (n.n.) 
hopping at half-filling, i.e., the 2D version of the 
well-known Su-Schrieffer-Heeger (SSH) model \cite{SSH}. 
In this case the FS consists of 
parallel straight lines: $|k_x|+|k_y|=\pi$ as illustrated in Fig. 1 by the 
solid line. One of the Fermi lines may be completely moved to another 
by a translational vector $Q=(\pi,\pi)$, i.e., the FS is perfectly 
nested with nesting vector $Q$. The Peierls instability for this model
was theoretically studied one decade ago in connection to high $T_c$ 
superconductors \cite{Tang,Mazumdar}. Unlike in one dimension, 
there are several possible 
alternation patterns for the lattice distortion and the corresponding bond 
hopping, as discussed by Tang and Hirsch \cite{Tang}. In Fig. \ref{Fig:Pattern}
two possible dimerization patterns are shown. Both of them correspond to 
phonons with wave vector $(\pi,\pi)$, which is exactly $Q$ so that
the reduced Brillouin zone boundary after distortion perfectly meets
the original FS. The 
difference between them is that for case (a) the dimerization is in both
directions, while it is only in one direction for case (b). It was
found that even arbitrarily small lattice elastic strength 
will induce a lattice distortion into case (a) or (b), i.e, 
the Peierls instability is sure to occur even for this 2D model 
\cite{Tang,Mazumdar}. On the other hand, however, it
may be noticed that the present perfect nesting of the FS may be easily 
broken, for example, by introducing next nearest-neighbour (n.n.n.) hopping 
which is often not negligible. Then the following problem naturally 
arises: does the above Peierls instability still 
survive the imperfect nesting of the FS?

Actually, for those quasi-2D materials which 
show a Peierls instability perfect nesting of their Fermi surfaces is never 
present, but an approximate, so called hidden nesting exists \cite{Whangbo}. 
Also, it may be reasonable to expect that the Peierls instability will be 
suppressed if the FS is so far away from nesting that even no 
hidden nesting is present. As far as real materials are concerned
the shapes of the Fermi surfaces are obtained by band structure calculations 
and may be often rather complicated. Nevertheless, 
the tight-binding model with the n.n. and n.n.n. 
hopping should already be sufficient to simulate an essential property 
of real Fermi surfaces: whether they are nesting or not. 
Thus it is necessary to clarify how sensitive the Peierls instability is 
to the deviations of the FS from perfect nesting which is controlled 
by n.n.n. hopping. This is the topic addressed in this paper. 
A similar problem was studied previously by Lin et al. \cite{Lin}, however,
their study was only limited to the dimerization pattern (a). 
(Actually this pattern is not favorable in a large region of $t'$ as will be 
seen later.) In addition, the main result Fig. 7 in their work was not 
convincingly presented. We will include the two possible patterns (a) and (b) 
and address the unexpected competition between them.

\begin{figure}
\epsfxsize=6.5cm
\epsfysize=6cm
\centerline{\epsffile{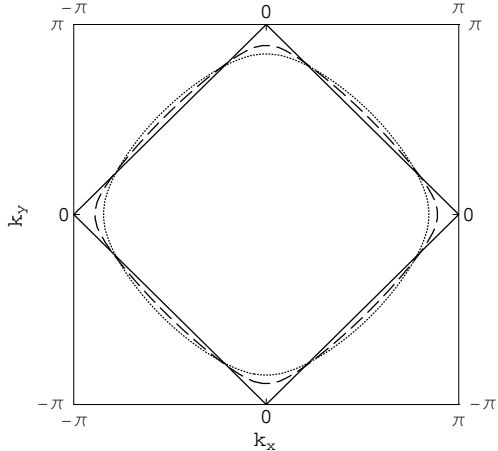}}
\medskip
\caption{The Fermi surfaces for the 2D tight-binding model without 
dimerization at half-filling. The solid line has $t'=0$, the dashed line has
$t'=0.1$, and the dotted line has $t'=0.2$.}
\label{Fig:FS}
\end{figure}

\begin{figure}
\epsfxsize=12cm
\epsfysize=6cm
\centerline{\epsffile{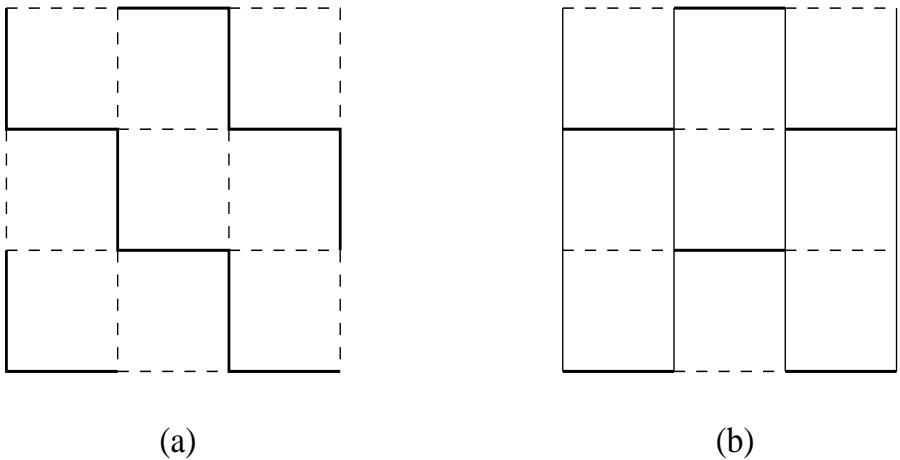}}
\medskip
\caption{The lattice distortion patterns (a) and (b). In the figure 
a thick solid line 
corresponds to a strong bond with hopping integral $t(1+\delta)$, a dashed 
line corresponds to a weak bond with hopping integral $t(1-\delta)$, and a 
thin solid line corresponds to a normal bond with hopping integral $t$. Both 
patterns correspond to phonons with wave vector $(\pi,\pi)$. The dimerization 
is along two axes for case (a), while only along $x$ axis for case (b).}
\label{Fig:Pattern}
\end{figure}


\section{FS nesting and Peierls instability}

We begin with the following Hamiltonian based on a square lattice with 
a half-filled band:
\begin{eqnarray}
H & = & -t\sum_{i,j,\sigma} [1+\alpha (u_{i,j}^x-u_{i+1,j}^x)]
  (c_{i,j,\sigma}^{\dagger}c_{i+1,j,\sigma}+{\rm h.c.})
-t\sum_{i,j,\sigma} [1+\alpha (u_{i,j}^y-u_{i,j+1}^y)]
  (c_{i,j,\sigma}^{\dagger}c_{i,j+1,\sigma}+{\rm h.c.}) \nonumber\\
& & -t'\sum_{i,j,\sigma}(c_{i,j,\sigma}^{\dagger}c_{i+1,j+1,\sigma}+
  c_{i,j,\sigma}^{\dagger}c_{i+1,j-1,\sigma}+ {\rm h.c.})+
{K\over 2} \sum_{i,j}[(u_{i,j}^x-u_{i+1,j}^x)^2+(u_{i,j}^y-u_{i,j+1}^y)^2]\ ,
\label{eq:H}
\end{eqnarray}
where $c_{i,j,\sigma}^{\dagger} (c_{i,j,\sigma})$ denotes the creation 
(annihilation) operator for an electron at site $(i,j)$ with spin $\sigma$
($i$ denotes $x$ coordinate and $j$ denotes $y$ coordinate),
$u_{i,j}^{x/y}$ represents the displacement component of site $(i,j)$ 
in $x/y$ direction,
$t,\ t'$ are n.n and n.n.n. hopping parameters and $\alpha$ is the 
electron-lattice coupling constant. The last term above describes the lattice 
elastic potential energy with $K$ the elastic constant. The lattice kinetic 
energy is omitted here since we do not study the dynamic behavior of phonons.
For the lattice distortion patterns investigated here, the distance between 
n.n.n. sites remains unchanged, therefore no dimerization of $t'$ is 
considered in Hamiltonian (\ref{eq:H}).

The lattice distortion, as shown in Fig. 2 having the wave vector 
$Q$, may be explicitly introduced as  
$$u_{i,j}^x-u_{i+1,j}^x=(-1)^{i+j}u,\ \  
  u_{i,j}^y-u_{i,j+1}^y=(-1)^{i+j}u
$$
for case (a) and
$$u_{i,j}^x-u_{i+1,j}^x=(-1)^{i+j}u,\ \  
  u_{i,j}^y-u_{i,j+1}^y=0\ \ \ \ \ 
$$
for case (b), where $u$ is the amplitude of dimerization determined by 
minimization of the ground state energy. For convenience, two dimensionless
parameters are defined as follows: the dimerization 
amplitude $\delta=\alpha u$ and the electron-lattice coupling 
constant $\eta= \alpha^2 t/K$. The n.n. hopping integral $t$ is taken as 
the energy unit.

Before proceeding, we would like to mention that two further possible lattice 
distortion patterns as discussed in previous works on the 2D Peierls 
instability, which correspond to wave vector $(\pi,0)$ 
and/or $(0,\pi)$ \cite{Tang}, are excluded in 
our study. We omit them because at $t'=0$ their energy gains are much smaller 
than those for the patterns considered here \cite{Tang}. (They are
possibly favorable only when a large Hubbard $U$ term is switched on 
\cite{Tang}.) This is physically understandable: they have a different wave 
vector from the nesting one. Very recently, a more 
complex lattice distortion pattern with incommensurate structure
was studied by Ono and Hamano in the case of $t'=0$ \cite{Ono}. 
However it will not be included here since it is not unique \cite{Ono}. 
The patterns considered in Fig. 2 may be regarded as 
typical structures for the study of the Peierls instability in two dimensions. 
In the following the Peierls instabilities for $t'=0$ and
$t'\neq 0$ are discussed respectively.

\subsection{$t'=0$}

For completeness, we first reproduce the results for the perfect nesting
case $t'=0$, which are straightforward but helpful. In momentum space 
the electronic spectra for case (a) and (b) are respectively written as 
\begin{eqnarray}
\varepsilon_{{\bf k},a}^{\pm} & = & \pm 2\sqrt{(\cos k_x+\cos k_y)^2+
\delta^2(\sin k_x+\sin k_y)^2}\ , \nonumber\\
\varepsilon_{{\bf k},b}^{\pm} & = & \pm 2\sqrt{(\cos k_x+\cos k_y)^2+
\delta^2\sin^2 k_x}\ .
\end{eqnarray}
In the ground state only the lower band is fully occupied for each case 
(i.e., the chemical potential is zero). Then the ground state energy is 
given by
$$
E=2\sum_{\bf k} \varepsilon_{{\bf k},a}^-+N\delta^2/\eta
$$
for case (a) and 
$$
E=2\sum_{\bf k} \varepsilon_{{\bf k},b}^-+N\delta^2/2\eta
$$
for case (b), where the summation is over the wave vector 
${\bf k}=(k_x,k_y)$ in the 
Brillouin zone $-\pi <k_x\pm k_y \le \pi$ and $N$ is the total number of 
lattice sites. The factor $2$ in front of the summation is due to spin 
degeneracy. Obviously the increase of elastic energy for case (a) is
two times of that for case (b).

\begin{figure}
\epsfxsize=8cm
\epsfysize=5.5cm
\centerline{\epsffile{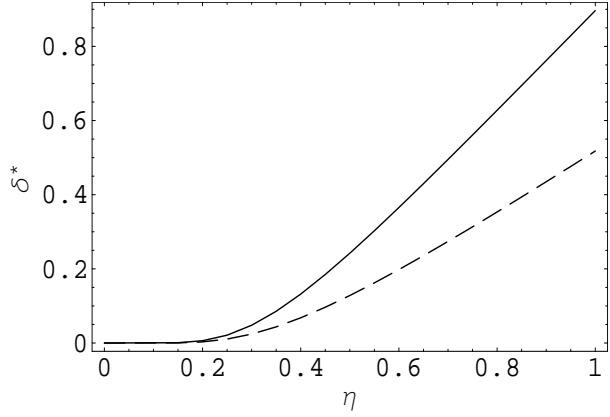}}
\medskip
\caption{The optimal value $\delta^*$ as a function of $\eta$ for $t'=0$.
The dashed line is for case (a) and the solid line is for case (b).}
\label{Fig:Del_td0}
\end{figure}

\begin{figure}
\epsfxsize=8.5cm
\epsfysize=5.5cm
\centerline{\epsffile{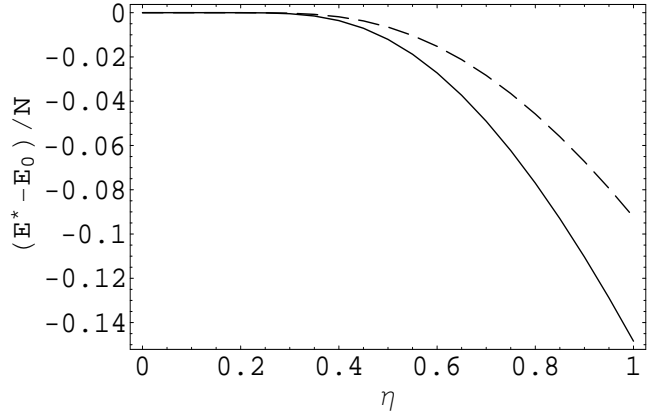}}
\medskip
\caption{The energy gain of the dimerized state: $E^*$ at $\delta=\delta^*$ 
minus $E_0$ at $\delta=0$ as a function of $\eta$, corresponding to 
Fig. \ref{Fig:Del_td0}.}
\label{Fig:E_td0}
\end{figure}

The results are shown in Figs. \ref{Fig:Del_td0} and 
\ref{Fig:E_td0}. 
In Fig. \ref{Fig:Del_td0} the optimal dimerization parameter $\delta^*$ 
with lowest energy vs. the electron-lattice coupling $\eta$ is plotted, 
and Fig. \ref{Fig:E_td0} 
gives the corresponding ground state energies $E^*$ at these $\delta^*$ values 
(with respect to the energy of the
undimerized lattice $E_0$ which is equal to $-16/\pi^2$ times $N$).
From these figures it can be seen that the Peierls instability takes place
as long as the electron-lattice coupling is non-zero for either case (a)
or (b). Although the magnitude of $\delta^*$ is small when 
$\eta \rightarrow 0$, it is proven to be finite, with an exponential 
dependence of $\exp (-c/\eta)$ on $\eta$ ($c$ is a constant). 
This confirms the related statement made in the Introduction. 
Moreover, it is seen from Fig. \ref{Fig:E_td0} that the energy $E^*$ for 
case (b) (solid line) is always lower than the corresponding one 
for case (a) (dashed line) in the full region of finite $\eta$, which means 
that the case (b) is the more favorable dimerization pattern for $t'=0$. 
Unfortunately in their works, Mazumdar 
as well as Tang and Hirsch incorrectly prefered case (a), 
see Ref. \cite{Mazumdar}. A similar conclusion as ours was reached 
by Ono and Hamano \cite{Ono}.

\subsection{$t'\neq 0$}

For $t'\neq 0$ the perfect nesting of the FS is broken. 
In Fig. \ref{Fig:FS} the Fermi surfaces for the undimerized lattice are 
plotted for several $t'$ values. One may see that the bigger $t'$ is 
the farther the FS deviates from perfect nesting.

Still the Hamiltonian can be easily diagonalized, but the final results are
quite nontrivial. Now the electronic spectra become
\begin{eqnarray}
\varepsilon_{{\bf k},a}^{\pm} & = & -4t'\cos k_x \cos k_y
\pm 2\sqrt{(\cos k_x+\cos k_y)^2+\delta^2(\sin k_x+\sin k_y)^2}\ , \nonumber\\
\varepsilon_{{\bf k},b}^{\pm} & = & -4t'\cos k_x \cos k_y
 \pm 2\sqrt{(\cos k_x+\cos k_y)^2+\delta^2\sin^2 k_x}\ ,
\label{eq:spm}
\end{eqnarray}
and the ground state energy is
$$
E=2\sum_{\varepsilon_{{\bf k},a}^{\pm}\le \mu_a} 
\varepsilon_{{\bf k},a}^{\pm}+N\delta^2/\eta
$$
for case (a) and 
$$
E=2\sum_{\varepsilon_{{\bf k},b}^{\pm}\le \mu_b} 
\varepsilon_{{\bf k},b}^{\pm}+N\delta^2/2\eta
$$
for case (b), where $\mu _a$ and $\mu _b$ are chemical potentials for 
case (a) and (b), respectively. 
By solving the equation $\partial E/\partial \delta =0$ we may single out 
the optimal value $\delta^*$ for each $t'$ with fixed $\eta$. In the 
following a typical value $\eta =0.5$ is adopted \cite{Lin,Ono}. 

\begin{figure}
\epsfxsize=8.5cm
\epsfysize=5.5cm
\centerline{\epsffile{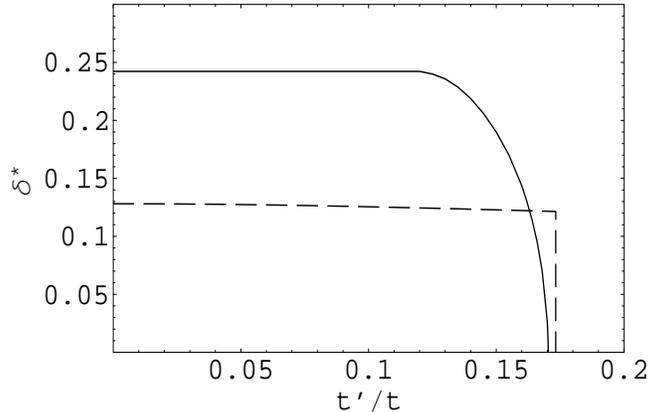}}
\medskip
\caption{The optimal value $\delta^*$ as a function of $t'$ for
$\eta=0.5$. The dashed line is for case (a) and the solid line is for 
case (b).}
\label{Fig:Del_eta05}
\end{figure}

The results for $\delta^*$ are shown in Fig. \ref{Fig:Del_eta05}. 
It is clearly seen that $\delta^*$ goes to zero with 
increase of $t'$ in both cases, which means that
the Peierls instability is suppressed at some $t'$, as expected. 
And more interestingly, the
details for both cases are different. For case (a) $\delta^*$ first
decreases weakly with increase of $t'$, but at some critical 
$t'_{c,1} \simeq 0.1733$, it drops suddenly to zero, showing a 
first-order transition. On the other hand, for case (b) $\delta^*$ 
first retains its $t'=0$ value, and
after $t'$ is beyond about $0.12$ it begins to decrease gradually and 
approaches zero smoothly at $t'_{c,2}\simeq 0.1704$ --- the transition is
of second-order. The value $t'_{c,2}$ is close to, but different from 
$t'_{c,1}$.

\begin{figure}
\epsfxsize=9.5cm
\epsfysize=6cm
\centerline{\epsffile{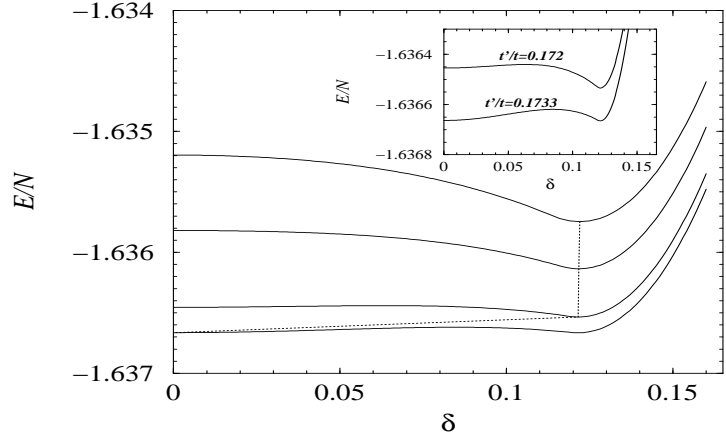}}
\medskip
\caption{The ground state energy per site as a function of $\delta$ for 
several $t'$ values for case (a). The solid curves from up to down 
correspond to $t'=0.164,\ 0.168,\ 0.172,\ 0.1733$, respectively, 
and the dotted line
connects the global minima of them. The curves for $t'=0.172$ and $t'=0.1733$ 
are enlarged in the inset, where the initial increase in each curve
is shown.}
\label{Fig:Es}
\end{figure}

\begin{figure}
\epsfxsize=9.5cm
\epsfysize=6cm
\centerline{\epsffile{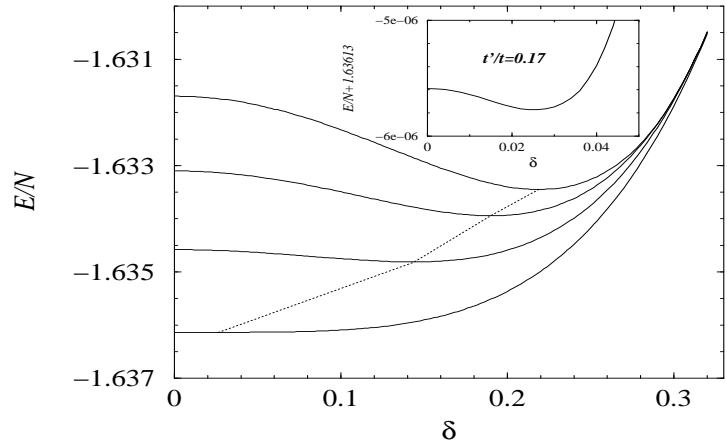}}
\medskip
\caption{Same as Fig. \ref{Fig:Es} for case (b). 
The curves from up to down correspond to $t'=0.14,\ 0.15,\ 0.16,\ 0.17$, 
respectively. In the inset the curve for $t'=0.17$ is shown 
in the small $\delta$ region.}
\label{Fig:Ea}
\end{figure}

To see the above transitions more clearly, we plot in Figs. \ref{Fig:Es} 
and \ref{Fig:Ea} the dependence
of the ground state energy on the dimerization parameter $\delta$ for several 
different $t'$ values for case (a) and (b), respectively. For case (a) it is
seen that when $t'$ approaches the critical value $t'_{c,1}$, 
the energy $E$ first increases with $\delta$ and then decreases towards 
a minimum, as shown in the inset of Fig. \ref{Fig:Es}. Thus this minimum 
may be only a local one. It is expected that at $t'_{c,1}$ the energy $E$ at 
such a minimum is the same as that at $\delta=0$, see the inset. 
And then once $t'$ is beyond $t'_{c,1}$ the $\delta$ value with absolutely 
lowest energy should be taken zero, i.e., $\delta^*=0$. So a first-order 
transition arises. On the other hand, for case (b) the energy $E$ always 
decreases first with increase of $\delta$ until a minimum is reached. 
So this minimum is actually a global one. It shifts continuously towards 
zero with increase of $t'$, which explains the second-order transition.

It deserves to point out the physical reason for the constant value of 
$\delta^*$ in the small $t'$ region (about $t'< 0.12$) for case (b). Actually,
in this region the two energy bands $\varepsilon_{{\bf k},b}^{\pm}$ do not 
overlap when one takes relatively large $\delta$ value like $\delta^*$, so
that the lower band $\varepsilon_{{\bf k},b}^{-}$ is fully occupied and 
the upper band $\varepsilon_{{\bf k},b}^{+}$ is empty in the ground state. 
Then the electronic energy is simply given by
$\sum_{\bf k}\varepsilon_{{\bf k},b}^{-}$, which is independent of
$t'$ due to $\sum_{\bf k}\cos k_x\cos k_y=0$. Therefore the ground state energy
for $t'\neq 0$ is the same as that for $t'=0$, so is the solution
$\delta^*$.

\begin{figure}
\epsfxsize=8.5cm
\epsfysize=5.5cm
\centerline{\epsffile{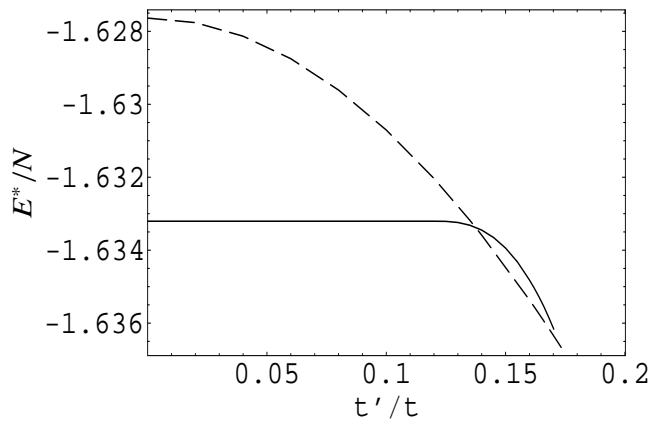}}
\medskip
\caption{The energy at $\delta=\delta^*$ as a function of $t'$ for 
$\eta=0.5$, corresponding to Fig. \ref{Fig:Del_eta05}.
Both curves stop at their respective critical points.}
\label{Fig:E_eta05}
\end{figure}

While so far the results of the Peierls instabilities for the two patterns 
have been given separately, it is now natural to think of the competition 
between them. The competition becomes evident when the respective lowest 
energies $E^*$ at different $t'$ values are compared, as shown in 
Fig. \ref{Fig:E_eta05}. It is interesting to see that
the energy $E^*$ for case (b) is first less than that for case (a), but later
becomes larger with increase of $t'$. Consequently, a transition
between these two dimerization patterns is predicted. To summarize, 
the evolution of the stable state for the system with increase of $t'$ 
may be described as follows: at $t'=0$ the dimerized state stabilizes 
due to the perfect FS 
nesting, and moreover the dimerization pattern takes form (b). This 
state remains stable until at some $t'$ value (which is about $0.13$ 
for $\eta=0.5$) it is replaced by the other dimerized state (a).
Finally, the dimerized state breaks down, i.e., the 
Peierls instability is suppressed at some critical value $t'_c$
which is equal to $t'_{c,1}$ as can be seen in Fig. \ref{Fig:E_eta05}. Thus the 
suppression is a first-order transition.

Above, only the results for $t'>0$ are presented. What is expected for
$t'<0$? Actually, taking $t'\rightarrow -t'$ reflects all the results as 
shown by Figs. \ref{Fig:Del_eta05} and \ref{Fig:E_eta05} about $t'=0$. 
This conclusion may be proven as follows. 
From Eq. (\ref{eq:spm}) one may get the relations between the 
electronic spectra $\varepsilon_{\bf k}^{\pm}$ and their 
corresponding ones ($\bar{\varepsilon}_{\bf k}^{\pm}$) 
after reflection $t'\rightarrow -t'$: 
$\bar{\varepsilon}_{\bf k}^{\pm}=-\varepsilon_{\bf k}^{\mp}$. 
(Because of the irrelevance of the proofs to the 
dimerization pattern the subscript which differentiates case (a) or (b) is 
omitted.) Then one obtains that the chemical potential $\bar{\mu}$
for the spectra $\bar{\varepsilon}_{\bf k}^{\pm}$ is just the negative of that 
for the $\varepsilon_{\bf k}^{\pm}$, i.e., $\bar{\mu}=-\mu$ at half-filling, 
due to the following equations: 
$$
\sum_{\varepsilon_{\bf k}^{\pm}<\mu} 1=\sum_{\varepsilon_{\bf k}^{\pm}>\mu} 1=
\sum_{-\varepsilon_{\bf k}^{\pm}<-\mu} 1=
\sum_{\bar{\varepsilon}_{\bf k}^{\pm}<-\mu} 1=N/2\ .
$$
In addition, one needs to notice that the summation for all energy levels 
$\sum_{\bf k} \varepsilon_{\bf k}^{\pm}=\sum_{\bf k} -8t'\cos k_x \cos k_y 
=0$. Finally the following relations hold:
$$
\sum_{\varepsilon_{\bf k}^{\pm}<\mu}\varepsilon_{\bf k}^{\pm}=
-\sum_{\varepsilon_{\bf k}^{\pm}>\mu}\varepsilon_{\bf k}^{\pm} =
-\sum_{-\varepsilon_{\bf k}^{\pm}<-\mu} \varepsilon_{\bf k}^{\pm}=
\sum_{\bar{\varepsilon}_{\bf k}^{\pm}<\bar{\mu}} \bar{\varepsilon}_{\bf k}^{\pm}\ .
$$
Note that the first and the last terms in the above formulas are nothing 
but the electronic parts of the ground state energies for $t'$ and $-t'$, 
respectively. They are proven to be the same, so are all the subsequent 
results derived from them.

\begin{figure}
\epsfxsize=8.5cm
\epsfysize=5.5cm
\centerline{\epsffile{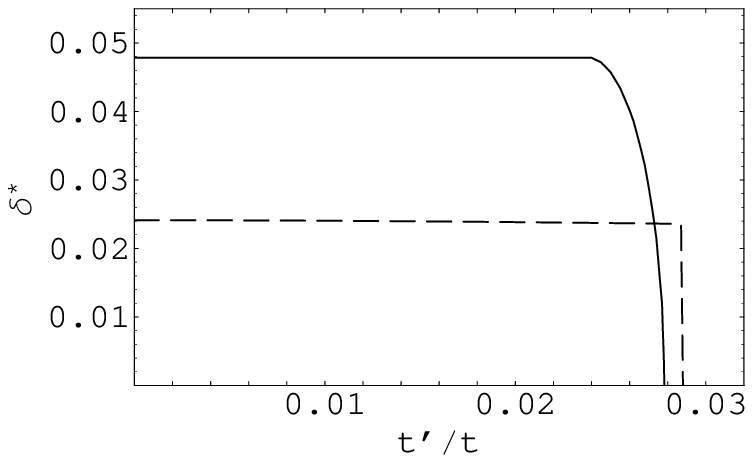}}
\medskip
\caption{The same as Fig. \ref{Fig:Del_eta05} but with $\eta=0.3$.}
\label{Fig:Del_eta03}
\end{figure}

\begin{figure}
\epsfxsize=8.5cm
\epsfysize=5.5cm
\centerline{\epsffile{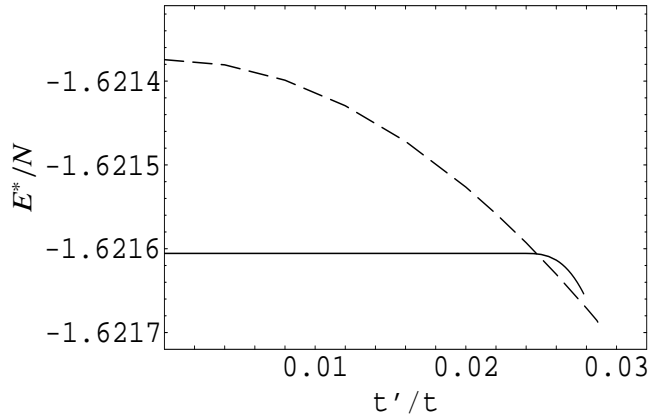}}
\medskip
\caption{The same as Fig. \ref{Fig:E_eta05} but with $\eta=0.3$.}
\label{Fig:E_eta03}
\end{figure}

Up to now all results have been shown for fixed $\eta$. In the last part of 
this section we will look at the role of $\eta$. 
Since the parameter $\eta$ only
appears in the elastic energy, i.e., it is irrelevant to the electronic part 
of the ground state energy, the qualitative properties of curves $E$ vs. 
$\delta$ as shown in Figs. \ref{Fig:Es} and \ref{Fig:Ea} are expected to be
retained, so are the above qualitative results for the Peierls instability. 
Quantitatively one may think, the smaller 
the electron-lattice coupling $\eta$ is (or equivalently the larger the 
elastic strength $K$ is), the smaller the critical value $t'_c$ for the
suppression of the Peierls instability should be. As a check, 
the case for $\eta=0.3$ is calculated. The results are shown in 
Figs. \ref{Fig:Del_eta03} and \ref{Fig:E_eta03}, which confirm the above 
predictions.

\section{Discussion}

In the previous section the Peierls instability for two dimerization 
patterns has been carefully studied in its dependence on $t'$.
The results are instructive. It is now known that low-dimensional metals
show two types of electronic instabilities: either a Peierls instability or
a superconducting one. The Peierls instability may prevent some metals
from entering a superconducting state. For example, in the transition metal
bronzes, the Peierls instability is the dominant mechanism and 
superconductivity was only found in LiMo$_6$O$_{17}$ \cite{Schlenker}. 
Our results suggest that the Peierls
instability may be suppressed by some way of increasing imperfect nesting, for
example, by applying pressure to enhance the n.n.n. hopping. Moreover,
the possible transition between different dimerized states is reminiscent of 
the experimental results in monophosphate tungsten bronzes 
like P$_4$W$_{12}$O$_{44}$, where double 
Peierls transitions occur with change of temperature \cite{Schlenker,Wang}. 
The competition between two dimerization patterns at finite 
temperatures is believed to be interesting and will be left 
for future investigation. 

A further problem is the consideration of effects of electron correlations, 
e.g., the Hubbard $U$ term which is not included in this work. 
For $t'=0$ the problem has been 
studied by Tang and Hirsch by numerical calculations \cite{Tang,Mazumdar}. 
It was found that the on-site 
Coulomb interaction weakens the dimerization in two dimensions as soon as
$U$ is present \cite{Mazumdar}. The possible explanation
is that the $U$ term favors the appearance of antiferromagnetic (AF) spin 
order in 2D half-filled Hubbard model, while the dimerization stabilizes 
local spin singlets which are unfavorable for strong $U$. 
Above, we showed that $t'$ also suppresses dimerization. 
Does this indicate that the 
simultaneous presence of both $U$ and $t'$ will speed the suppression of the 
dimerization? We leave this issue for future work and only give some clues 
here from previous analysis in the large $U$ limit 
(for the moment the concrete dimerization pattern is not considered).
In this limit we are facing a problem of localized electrons
interacting via an effective n.n. exchange $J$ ($\sim t^2/U$) and 
n.n.n. exchange $J'$ ($\sim t'^2/U$). The Peierls system then transforms into 
the corresponding spin-Peierls (SP) system. The 2D (or quasi-1D) spin-Peierls 
instability without n.n.n. exchange $J'$ was studied by some authors 
\cite{Inagaki,Yuan}. It was found that the SP transition does not
spontaneously occur unless the so called spin-lattice coupling (analogous to 
$\eta$ here) exceeds a threshold. On the other hand, it was known that 
the n.n.n. exchange $J'$ will frustrate the AF order for the 2D spin system. 
So one may assume that the additional inclusion of $J'$ will be 
favorable to the formation of a SP state. Thus it is not unreasonable to 
believe that the effects of $t'$ and $U$ on the Peierls instability
may cancel in part when acting simultaneously,
although each of them separately tends to suppress it.

\section{Conclusion}
In conclusion, a 2D tight-binding model with n.n. hopping $t$ and n.n.n. 
hopping $t'$ is used to study the effect of imperfect FS nesting on the 
Peierls instability of the ground state. Two possible dimerization patterns 
corresponding to a phonon vector $(\pi, \pi)$ are considered as case (a) 
and (b). It is found that 
the Peierls instability will be suppressed with an increase of $t'$ 
which characterizes the deviation from perfect nesting. The details for the 
two cases are different: for case (a) the suppression is a first-order 
transition while for case (b) it is of second-order. Also a transition 
between the two dimerized states is investigated.

\medskip

We acknowledge the financial support by the Deutsche Forschungsgemeinschaft
through SFB 484 and the BMBF 13N6918/1. Q. Yuan also acknowledges the partial
support by the Chinese NSF.


\begin{references}
\bibitem{Gruner} G. Gr\"uner, {\it Density Waves in Solids} (Addison-Wesley, 
 Redwood City, 1994). 
\bibitem{Farges} J. P. Farges (Ed.), {\it Organic Conductors} (Marcel Dekker, 
 New York, 1994).
\bibitem{Schlenk} C. Schlenker, J. Dumas, M. Greenblatt 
 and S. van Smaalen (Eds.), {\it Physics and Chemistry of Low-Dimensional 
 Inorganic Conductors}, NATO ASI Series B: Physics Vol. 354 
 (Plenum, New York, 1996). 
\bibitem{Dumas} J. Dumas and C. Schlenker, Int. J. Mod. Phys. B {\bf 7}, 4045 
 (1993).
\bibitem{Qin} X. K. Qin, J. Shi, H. Y. Gong, M. L. Tian, J. Y. Wei,
 H. Chen and D. C. Tian, Phys. Rev. B {\bf 53}, 15538 (1996); R. Xiong, 
 Q. M. Xiao, J. Shi, H. L. Liu, W. F. Tang, M. L. Tian and D. C. Tian,
 Mod. Phys. Lett. B {\bf 14}, 345 (2000).
\bibitem{Schlenker} C. Schlenker, C. Hess, C. Le Touze and J. Dumas,
 J. Phys. I {\bf 6}, 2061 (1996).
\bibitem{Wang} E. Wang, M. Greenblatt, I. E. Rachidi,
 E. Canadell, M. H. Whangbo and S. Vadlamannati, 
 Phys. Rev. B {\bf 39}, 12969 (1989); S. Drouard, D. Groult, J. Dumas,
 R. Buder and C. Schlenker, Eur. Phys. J. B {\bf 16}, 593 (2000).
\bibitem{Zheng} For the study on the effects of quantum lattice 
 fluctuations, see e.g., H. Zheng, Phys. Rev. B {\bf 50}, 6717 (1994).
\bibitem{SSH} A. J. Heeger, S. Kivelson, J. R. Schrieffer and W. P. Su,
 Rev. Mod. Phys., {\bf 60}, 781 (1988).
\bibitem{Tang} S. Tang and J. E. Hirsch, Phys. Rev. B {\bf 37}, 9546 (1988).
\bibitem{Mazumdar} S. Mazumdar, Phys. Rev. B {\bf 39}, 12324 (1989); S. Tang 
 and J. E. Hirsch, Phys. Rev. B {\bf 39}, 12327 (1989).
\bibitem{Whangbo} M. H. Whangbo, E. Canadell, P. Foury and J. P. Pouget,
  Science, {\bf 252}, 96 (1991).
\bibitem{Lin} F. Lin, X. B. Chen, R. T. Fu, X. Sun and Y. Kawazoe,
  Phys. Stat. Sol. (b) {\bf 206}, 559 (1998). 
\bibitem{Ono} Y. Ono and T. Hamano, J. Phys. Soc. Jpn. {\bf 69}, 1769 (2000).
\bibitem{Inagaki} S. Inagaki and H. Fukuyama, J. Phys. Soc. Jpn. {\bf 52}, 
 3620 (1983).
\bibitem{Yuan} Q. S. Yuan, Y. M. Zhang and H. Chen, cond-mat/9911119. 
\end{references}
\end{document}